\documentclass[twocolumn,showpacs,preprintnumbers,amsmath,amssymb,prb,aps]{revtex4-2} 
\usepackage{epsfig}
\usepackage{epstopdf}
\usepackage{multirow}
\usepackage{graphicx}
\usepackage{dcolumn}
\usepackage{bm}
\usepackage{textcomp}
\usepackage{array}
\usepackage{csquotes}
\usepackage{subcaption}
\usepackage{booktabs}
\usepackage{amsmath}
\usepackage{hyperref}
\hypersetup{
    colorlinks=true,
    linkcolor=blue,
    filecolor=magenta,      
    urlcolor=blue,
    pdftitle={},
    }
    
\begin{document}
\title{Investigating the effect of electronic correlation on transport properties and phononic states of Vanadium}
\author{Prakash Pandey$^{1}$}
\altaffiliation{ \url{prakashpandey6215@gmail.com}}
\author{Vivek Pandey$^{1}$}
\author{Sudhir K. Pandey$^{2}$}
\altaffiliation{ \url{sudhir@iitmandi.ac.in}}
\affiliation{$^{1}$School of Physical Sciences, Indian Institute of Technology Mandi, Kamand - 175075, India\\$^{2}$School of Mechanical and Materials Engineering, Indian Institute of Technology Mandi, Kamand - 175075, India}

\date{\today}

\begin{abstract}
In the present work, we have tried to investigate the importance of electronic correlation on transport properties and phononic states of Vanadium (V). The temperature-dependent electrical resistivity ($\rho$) and electronic part of thermal conductivity ($\kappa_e$) due to electron-electron interactions (EEIs) and electron-phonon interactions (EPIs) are computed. The values of $\rho$ due to EEIs are found to be extremely small in comparison to $\rho$ due to EPIs. For instance, at 300 K, the calculated value of $\rho$ due to EEIs (EPIs) is $\sim$ 0.859$\times10^{-3}$ ($\sim$ 0.20) $\mu\Omega$m. The magnitudes of $\kappa_e$ due to EPIs are found to be in good agreement with the experimental results. These observations indicate the negligible importance of EEIs to these quantities for V. However, at 300 K, the value of Seebeck coefficient ($S$) at DFT+DMFT level ($\sim$ -0.547 $\mu$VK$^{-1}$) is found to be entirely different than at DFT level ($\sim 7.401$ $\mu$VK$^{-1}$). Also, the DFT+DMFT value of $S$ at 300 K is in good match with the available experimental data (-1.06 $\mu$VK$^{-1}$\cite{Okram}, 1.0 $\mu$VK$^{-1}$\cite{Mackintosh}). Apart from this, the study of phononic states at DFT and DFT+DMFT level is performed. The obtained phononic band structure and phonon DOS at DFT+DMFT differ to a good extent from that at DFT. The maximum energy of phononic state obtained at DFT (DFT+DMFT) is $\sim$ 33.83 ($\sim$ 35.15) meV, where the result of DFT+DMFT is obtained more closer to the experimental data (35.15\cite{Stewart}, 36.98\cite{Mozer} \& 41.77\cite{chernoplekov1963investigation} meV). These results highlight the importance of electronic correlation on $S$ \& phononic states of simple correlated V metal.
\end{abstract}

\maketitle

\section{Introduction} 
In the past few years, physicists have studied a lot about the effect of electronic correlation on various transport properties of materials\cite{PhysRevB.86.115418, Zhang2015, PhysRevB.56.R4317, PhysRevB.78.012404}. The strength of electronic correlation in a material depends upon $U/t$ ratio, where $U$ is on-site Coulomb interaction and $t$ is hopping integral. Here, $t$ is related to band width ($W$). Based on these parameters, materials are usually classified as weakly ($W\gg U$), moderately ($W\approx U$) and strongly correlated ($W\ll U$) systems\cite{annurev-matsci-070218-121825, RevModPhys.78.865}. A wide range of novel quantum phenomena arise due to the electronic correlation of $d$ and $f$ electrons. Some of these are superconductivity\cite{ashcroft}, topological properties\cite{PhysRevLett.110.096401}, spin and cluster-glasses\cite{RevModPhys.58.801}, Kondo effect\cite{Kouwenhoven_2001}, etc. Thus, the theoretical study of the effect of electronic correlation has generated great interest in the physics community\cite{PhysRevLett.104.106408, PhysRevB.103.L241117}. However, theoretical developments for studying these systems are still a challenging task. This is because the many-body Hamiltonian equations that define the system are not exactly solvable using the existing theoretical approach. In the case of correlated electron systems (CESs), dynamical mean field theory (DMFT)\cite{RevModPhys.68.13, PRBhaule} is successful in capturing the electronic correlation up to a great extent.

With more than fifty years of research, $3d$ transition metals have attracted tremendous interest in the scientific community for their fascinating behaviour\cite{PhysRev.134.A923, PhysRev.139.A1905}. Recent studies on Vanadium (V) categorise it as a moderately correlated electron system\cite{sihi2020detailed}. Thus, it would be a suitable candidate for studying the effect of electronic correlation on various transport properties and phononic states. Also, Belozerov \textit{et al.}\cite{PhysRevB.107.035116} have recently studied electronic correlation and its effect on magnetic properties of V using DFT+DMFT method. Their results indicate that V shows Pauli paramagnetism at low temperatures, while it follows the Curie-Weiss law at higher temperatures. Also, they reported the quasiparticle mass enhancement factor $\frac{m}{m^*}$ of $e_g$ state to be 1.4 at $U=2.3$ eV. Our result of $t_{2g}$ ($e_g$), \textit{i.e.} 1.14 (1.12) at $U=2.6$ eV using the DFT method is in a good agreement with their results. Furthermore, Savrasov \textit{et al.}\cite{Savrasov3} have studied the electrical resistivity ($\rho$) and electronic part of thermal conductivity ($\kappa_e$) of V due to electron-phonon interactions (EPIs) using density-functional theory (DFT) method. In this work, they have also compared their theoretical results with an experimental data set, which are in a good match\cite{Hellwege}. However, at present, there are several experimental results available corresponding to $\rho$\cite{chernoplekov1973change, westlake1967resistometric, pan1977investigation, Desai}. But these data differ from each other to a good extent.\cite{chernoplekov1973change, westlake1967resistometric, pan1977investigation, Desai}. So it becomes a dilemma as to which experimental data is more accurate. In such circumstances, even the theoretical data obtained at the DFT level is questionable. Thus, it becomes necessary to investigate the $\rho$ of V at DFT+DMFT level to verify whether the electronic correlation affects it or not. According to the Wiedemann-Franz law, $\kappa_e$ of a material is directly related to $\rho$\cite{ashcroft}. Thus, the effect of correlation on $\kappa_e$ is expected to be similar to $\rho$. Since, the value of $U$ for V is $2.6$ eV, which lies in the range of moderately CESs\cite{sihi2020detailed}. So, DFT may not truly capture the electronic correlation in this case. Sihi \textit{et al.}\cite{sihi2020detailed} have studied the photoemission spectra (PES) of V qualitatively using different theoretical approaches. In this work, it was shown that the qualitative features of PES, near the Fermi level, obtained through DFT and DFT+DMFT methods are in good agreement with the experimental data. However, the quantitative comparison of PES at both the levels is still lacking in their work. Furthermore, mere matching of PES does not guarantee that the features of states and their distributions around the Fermi energy will be same at DFT and DFT+DMFT levels. It is important to mention here that the transport properties are highly dependent on the states near the Fermi level. Thus, a comparative study of temperature-dependent transport properties of V using DFT and DFT+DMFT approach is very much needed to explore the importance of electronic correlation on these properties of the material.

Phonons play essential roles in deciding several material properties such as thermal and electronic transport, thermodynamics, and structural stability. The phonon calculations are usually performed for bulk materials through the DFT method, which provides good results for most of the time\cite{togo, PhysRevB.84.094302}. However, generally used exchange-correlation functionals such as LDA\cite{Perdew} and GGA\cite{PhysRevLett.77.3865} do not properly incorporate the effect of electronic correlation. While tackling these problems, DFT+DMFT generally seems to offer a potential path for a better understanding of the effect of electronic correlation. Recently, Khanal \textit{et al.}\cite{PhysRevB.102.241108} studied the phononic states of iron superconductor FeSe using the DFT+DMFT approach. Their work highlights the effect of electronic correlations on phonon frequencies. Also, in a recent work, Koçer \textit{et al.}\cite{PhysRevB.102.245104} have provided an efficient and general approach for the phonon calculations by using DFT+DMFT method. In their work, they have calculated the lattice dynamics properties of Fe, NiO, MnO, and SrVO$_3$ using DFT+DMFT approach. Furthermore, using DFT+DMFT approach, the phonon spectra of plutonium and iron were also studied by several authors\cite{science.1083428, PhysRevB.85.020401, leonov2014, PhysRevLett.120.187203}. A key question that we would like to address is whether or not the electronic correlation affects the phononic states of V. This study would be helpful in further exploring the effects of electronic correlation on properties of V.

\begin{figure*}\label{Fig.1} 
\centering
\includegraphics[width=0.45\linewidth, height=5.0cm]{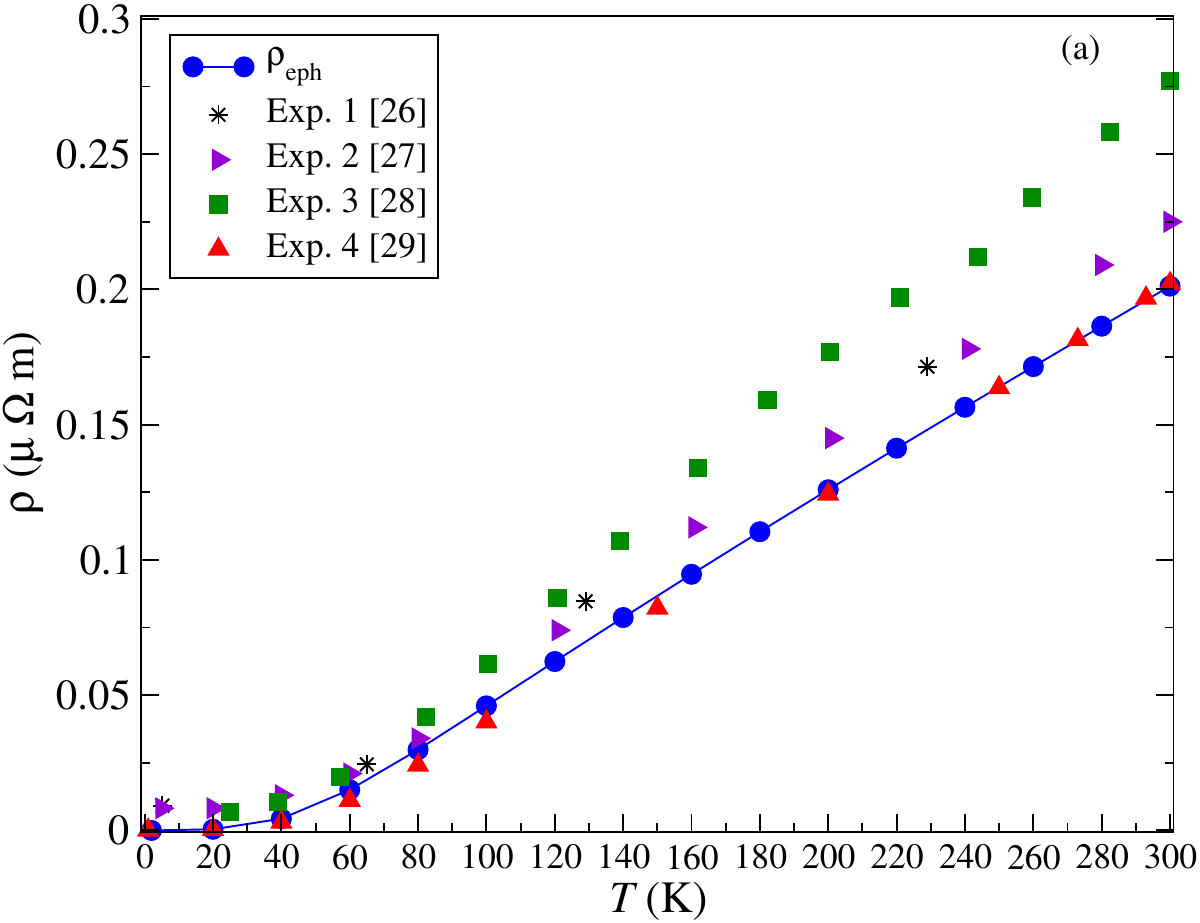}
\includegraphics[width=0.45\linewidth, height=5.0cm]{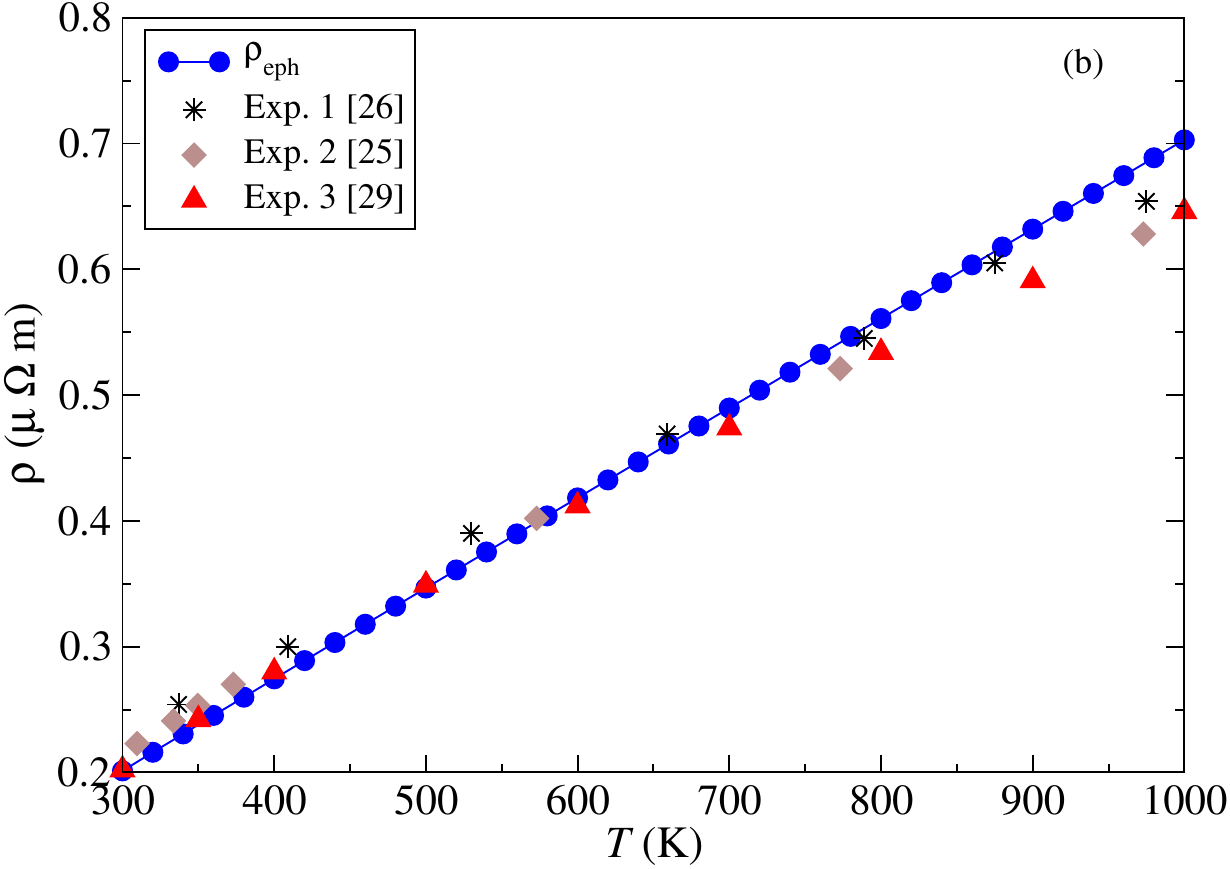}
\caption{\label{Fig.1}\small{Electrical resistivity ($\rho$) versus temperature ($T$) plot (a) from $0-300$ K (b) from $300-1000$ K. The connected blue data (splitted data-points) represents $\rho_{eph}$ (experimental results).}}
\end{figure*}
 
In the present work, an effort has been made to explore the extent to which electronic correlation affects the transport properties and phononic states of V. With this motivation, the contribution to the values of temperature-dependent $\rho$ and $\kappa_e$ due to electron-electron interactions (EEIs) and electron-phonon interactions (EPIs) are calculated. A comparative analysis of both these contributions to $\rho$ and $\kappa_e$ is done. In addition to this, the study of temperature-dependent Seebeck coefficient $S$ is carried out within the temperature range of $100-1000$ K using DFT+DMFT method. Also, the values of $S$ are computed at 300 K via DFT method. Apart from transport properties, the study of phononic states is carried out at DFT and DFT+DMFT levels. In this regard, the calculations of phonon dispersion curve and phonon DOS are performed using both DFT and DFT+DMFT methods. The comparative study of the result obtained using both the approaches highlights the importance of electronic correlation on the $S$ and phononic states of V.

\begin{table*}\label{tabb}
\caption{\label{tabb}
\small{Calculated values of electrical conductivity ($\sigma$), Seebeck coefficient ($S$) and electronic part of thermal conductivity ($\kappa_e^{ee}$) due to EEIs corresponding to different sizes of $k$-mesh at 300 K.}} 
\begin{ruledtabular}
\begin{tabular}{lccc}
\textrm{{$k$-mesh}}&
\textrm{{$\sigma$ (*$10^6\hspace{1mm}\Omega^{-1}$m$^{-1}$)}}&
\textrm{{$S$ ($\mu$VK$^{-1}$)}}& 
\textrm{{$\kappa_e^{ee}$ (Wm$^{-1}$K$^{-1})$}}\\

\colrule
 ($66\times66\times66$)    & 1256.55  & 12.975   & 6481.646 \\
 ($92\times92\times92$)    & 1114.97  & 16.827   & 7026.405 \\
 ($100\times100\times100$) & 1139.33  & -4.269   & 6453.981 \\
 ($120\times120\times120$) & 1144.65  & 0.205    & 7873.407 \\
 ($140\times140\times140$) & 1236.16  & 4.696    & 7849.347 \\
 ($150\times150\times150$) & 1453.71  & -10.459  & 8587.235 \\
 ($200\times200\times200$) & 1084.86  & 1.404    & 7225.235 \\
 ($220\times220\times220$) & 1158.14  & -1.108   & 7236.168 \\
 ($250\times250\times250$) & 1104.06  & 0.359    & 7395.98 \\
 ($300\times300\times300$) & 1163.52  & -0.547   & 7686.932 \\
\end{tabular}
\end{ruledtabular}
\end{table*}

\section{Computational details}
The density functional theory (DFT) based calculations are carried out using WIEN2k package\cite{blaha2020wien2k}. The non-magnetic calculations of electronic and phononic properties of Vanadium (V) are performed using full-potential linearized augmented plane wave (FP-LAPW) method. Local density approximation (LDA)\cite{Perdew} is used as exchange-correlation functional in these calculations. The space group and lattice constant for V are taken as \textit{Im$\bar{3}$m} and 3.024 \AA, respectively\cite{sihi2020detailed}. Furthermore, the Wyckoff position and the muffin-tin sphere radius for V atom are taken as $(0, 0, 0)$ and 2.0 Bohr, respectively. The self-consistent ground state energy calculations are performed over a \textit{k}-mesh size of 10$\times$10$\times$10 with the charge convergence limit set to $10^{-4}$ Ry/Bohr. The temperature-dependent DFT+dynamical mean field theory (DMFT) computations are carried out using embedded DMFT functional (eDMFTF) code\cite{PRBhaule}, which is interfaced with WIEN2k package. The ground state energy is computed in a self-consistent way over a \textit{k}-mesh size of 66$\times$66$\times$66. For these calculations, the energy convergence limit and the charge convergence limit are set to $10^{-3}$ Ry/Bohr and $10^{-4}$ Ry/Bohr, respectively. The value of $U=2.6$ eV is taken from the literature\cite{sihi2020detailed} and the value of $J=0.61$ eV is calculated by Yukawa screening method\cite{PhysRevLett.115.256402}. The exact double counting method is used to avoid the double-counting issue\cite{PhysRevLett.115.196403}. The continuous time quantum Monte Carlo (CTQMC) method is used as an impurity solver in these calculations\cite{PhysRevB.75.155113}. This CTQMC method is implemented in the eDMFTF code\cite{PRBhaule}. Also, the maximum entropy analytical continuation method is used to obtain the spectral function in real axis\cite{PhysRevLett.115.196403}. Density of states (DOS) and eigenvalues are calculated over a dense \textit{k}-mesh size of 300$\times$300$\times$300. These eigenvalues are further used to calculate the electronic transport properties using TRACK code\cite{sihi2021track}, which is interfaced with eDMFTF code. In the DFT+DMFT calculations, the Fermi energy is found to be 7.9911 eV. Using this Fermi energy, the velocity matrix is calculated via optic module of WIEN2k package\cite{blaha2020wien2k} over a dense \textit{k}-mesh size of 300$\times$300$\times$300.

It is important to mention here that before computing the temperature-dependent transport properties, a \textit{k}-mesh convergence test has been performed over these properties at 300 K. Corresponding data is mentioned in Table \ref{tabb}. The DFT and DFT+DMFT based phonon calculations are carried out for a 2$\times$2$\times$2 supercell using PHONOPY package\cite{togo}. The DFT based force calculations are performed over a \textit{k}-mesh size of 12$\times$12$\times$12 using WIEN2k package with the force convergence limit set to $10^{-5}$ Ry/Bohr. Furthermore, DMFT based force calculations are performed at 300 K using eDMFTF code.
 
The calculations of transport properties due to electron-phonon interactions (EPIs) are carried out using Abinit code\cite{gonze2016recent}. These computations are performed using density functional perturbation theory (DFPT) method\cite{PhysRevB.43.7231}. The projector augmented wave (PAW)\cite{Bloachi} pseudopotential is adopted to obtain transport properties. These calculations are performed over a \textit{k}-mesh size of 42$\times$42$\times$42 and \textit{q}-mesh size of 6$\times$6$\times$6. The energy cutoff parameter is set to 30 Hartree.

\section{Results and Discussion}
\subsection{\label{sec:level2}Transport properties}
The electrical resistivity ($\rho$) is an important transport property that aids in determining the practical applications of materials. The calculations of temperature ($T$) dependent $\rho$ due to electron-phonon interactions (EPIs) are carried out within the temperature range of $0 - 1000$ K. The experimental results are available in two different temperature ranges ($0-300$ K and $300-1000$ K). So the calculated results are divided into two parts for the purpose of comparison with the available experimental data. Corresponding plots are shown in Fig(s). \ref{Fig.1}(a) and \ref{Fig.1}(b), respectively. In both the Fig(s)., it is seen that the values of $\rho$ increase with the rise in $T$. The magnitude of $\rho$ due to EPIs ($\rho_{eph}$) is obtained as $\sim$ 0.42$\times10^{-3}$ $\mu\Omega$m at 20 K, which gets raised to $\sim$ 0.20 $\mu\Omega$m at 300 K. Furthermore, the magnitude of $\rho_{eph}$ at 1000 K is found to be $\sim$ 0.7029 $\mu\Omega$m. The obtained results are also compared with the available experimental data\cite{Hellwege, chernoplekov1973change, westlake1967resistometric, pan1977investigation, Desai}. For a better understanding of the effect of electronic correlation on $\rho$ of the material, the temperature-dependent $\rho$ due to electron-electron interactions (EEIs) are also calculated using DFT+DMFT method. Corresponding results are mentioned in Table \ref{taba}. The magnitude of $\rho$ due to EEIs ($\rho_{ee}$) is obtained as $\sim$ 0.9436$\times10^{-3}$ $\mu\Omega$m at 100 K, which gets raised to $\sim$ 0.15$\times10^{-2}$ $\mu\Omega$m at 1000 K. It is seen from the table and Fig(s). \ref{Fig.1} that, within the given temperature range, at any value of $T$, the magnitude of $\rho_{ee}$ is even less than $0.5\%$ of the value of $\rho_{eph}$. The possible reason behind this may be the extremely large order of electron-electron lifetime ($\tau_{ee}$) in comparison to that of electron-phonon lifetime ($\tau_{eph}$). The calculated order of $\tau_{ee}$ and $\tau_{eph}$ are $10^{-8}$ and $10^{-14}$ seconds, respectively.

\begin{table}\label{taba}
\caption{\label{taba}%
\small{Calculated values of electrical resistivity ($\rho_{ee}$), electronic part of thermal conductivity ($\kappa_e^{ee}$) due to EEIs and Lorenz number $(L)$ at different temperatures ($T$).}}
\begin{ruledtabular}
\begin{tabular}{lcccc}
\textrm{{$T$ }}&
\textrm{{$\rho_{ee}$ }}&
\textrm{{$\kappa_e^{ee}$}} &
\textrm{{$L$ }}\\  
\textrm{{(K)}}&
\textrm{{ (*$10^{-2}\mu\Omega$m)}}&
\textrm{{(Wm$^{-1}$K$^{-1}$)}} &
\textrm{{ ($\mu$V$^{2}$K$^{-2}$)}}\\ 
      
\colrule
 100    & 0.09436    & 2728 & 0.026  \\
 200    & 0.0856     & 4360 & 0.019  \\
 300    & 0.0859     & 7687 & 0.022  \\
 400    & 0.0895     & 7690 & 0.017  \\
 500    & 0.0965     & 8889 & 0.017  \\
 600    & 0.1054     & 10245 & 0.018 \\
 700    & 0.1151     & 11706 & 0.019 \\
 800    & 0.1684     & 13158 & 0.028 \\
 900    & 0.1435     & 10065 & 0.016 \\
 1000   & 0.15       & 15869 & 0.024 \\
\end{tabular}
\end{ruledtabular}
\end{table} 

Apart from the slight deviations, the calculated values of $\rho$ within the temperature range $0-1000$ K, are found to be in a good match with the available experimental data in literature\cite{Hellwege, chernoplekov1973change, westlake1967resistometric, pan1977investigation, Desai}. Furthermore, Table \ref{taba} shows that the magnitudes of $\rho_{ee}$ increase with the rise in temperature, except for some values of $T$. The obtained value of $\rho_{ee}$ at 100 K is greater than the value of $\rho_{ee}$ at 200 K. This is possibly because in the DFT+DMFT calculations, eigenvalues are computed using maximum entropy method\cite{PhysRevLett.115.196403} which may not provide accurate results at lower values of $T$. As these eigenvalues are involved in calculating $\rho_{ee}$, high fluctuations are expected at small values of $T$. Apart from the EEIs, EPIs greatly affect the values of $\rho$ as seen in Fig. \ref{Fig.1}. As the phonons cause the scattering of electrons, the increase in number of phonons will enhance the scattering process. This will result in the increase in values of $\rho$ with the rise in $T$. The discussion of the effect of EPIs on $\rho$ suggests that the values of $\rho_{eph}$ must increase with the rise in $T$. The results obtained in Fig. \ref{Fig.1} are in accordance with this. As already discussed, the contribution of EEIs to the values of $\rho$ is extremely small in comparison to that of EPIs at any given $T$. This indicates that, within the given temperature range, the effect of EEIs is insignificant in deciding the value of $\rho$. This further suggests that electronic correlation is negligibly importance to the values of $\rho$ of V within the given temperature range.


The importance of electronic correlation on the electronic part of thermal conductivity ($\kappa_e$) of the material is also investigated. In light of this, the calculations of temperature-dependent $\kappa_e$ due to EPIs ($\kappa_{eph}$) are performed within the temperature range of  $20-300$ K and the results are depicted in Fig. \ref{Fig.wth}. It is seen from the figure that the values of $\kappa_{eph}$ decrease with the rise in $T$. The magnitude of $\kappa_{eph}$ is found to be $\sim$ 280.2 Wm$^{-1}$K$^{-1}$ at $20$ K which gets decreased to $\sim$ 35.9 Wm$^{-1}$K$^{-1}$ at $300$ K. The figure also shows the comparison between the obtained results and available experimental data\cite{Hellwege}. In addition to this, the thermal conductivity due to EEIs ($\kappa_e^{ee}$) are also calculated within the temperature range of $100-1000$ K. The obtained results are mentioned in Table \ref{taba}. It is found from the table that values of $\kappa_e^{ee}$ increase with the rise in $T$. The magnitude of $\kappa_e^{ee}$ is found to be $\sim$ 2728 Wm$^{-1}$K$^{-1}$ at $100$ K which gets increased to $\sim$ 15869 Wm$^{-1}$K$^{-1}$ at $1000$ K.

\begin{figure}\label{Fig.wth}
\includegraphics[width=0.70\linewidth, height=5.0cm]{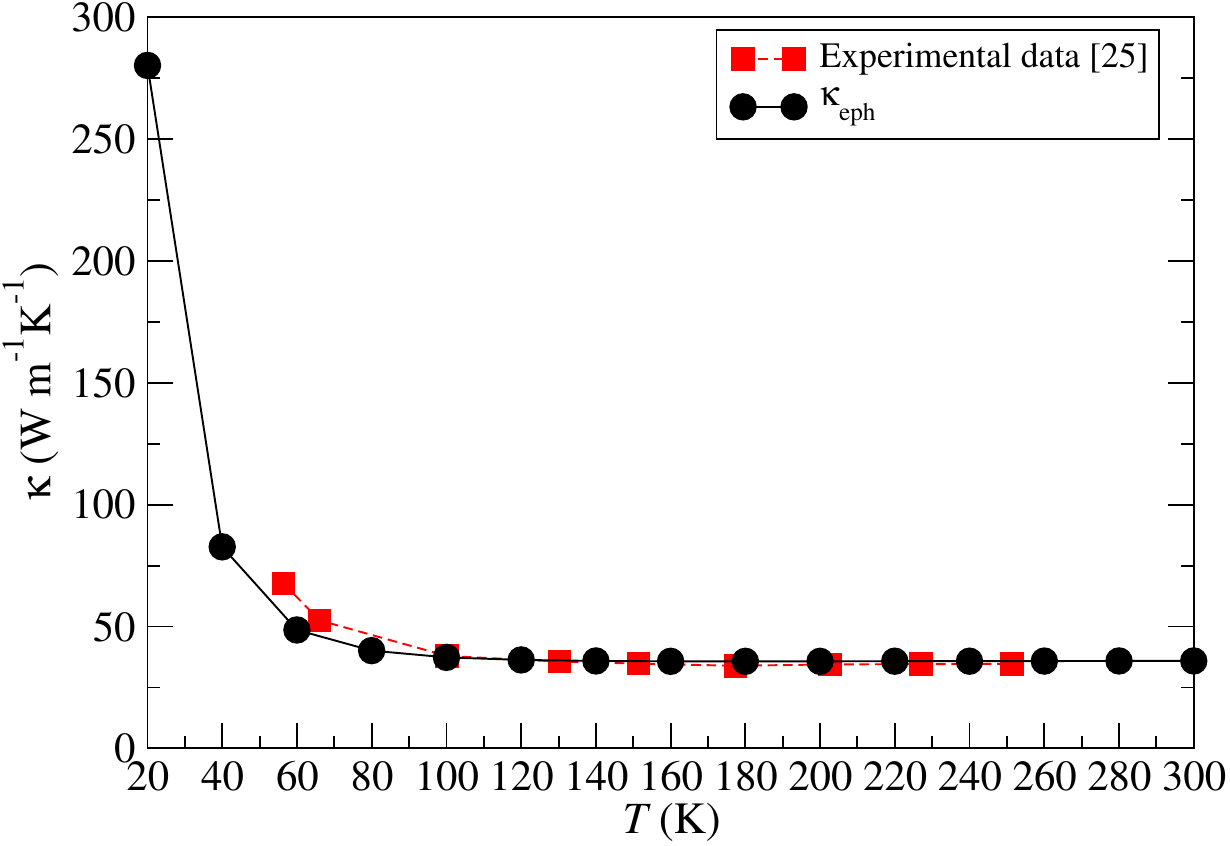} 
\caption{\label{Fig.wth}\small{Thermal conductivity ($\kappa$) versus temperature ($T$) plot. The black (red) coloured curve represents $\kappa_{eph}$ (experimental) data.}}
\end{figure}

\begin{figure}\label{Fig.self_energy}
\includegraphics[width=0.7\linewidth, height=5.0cm]{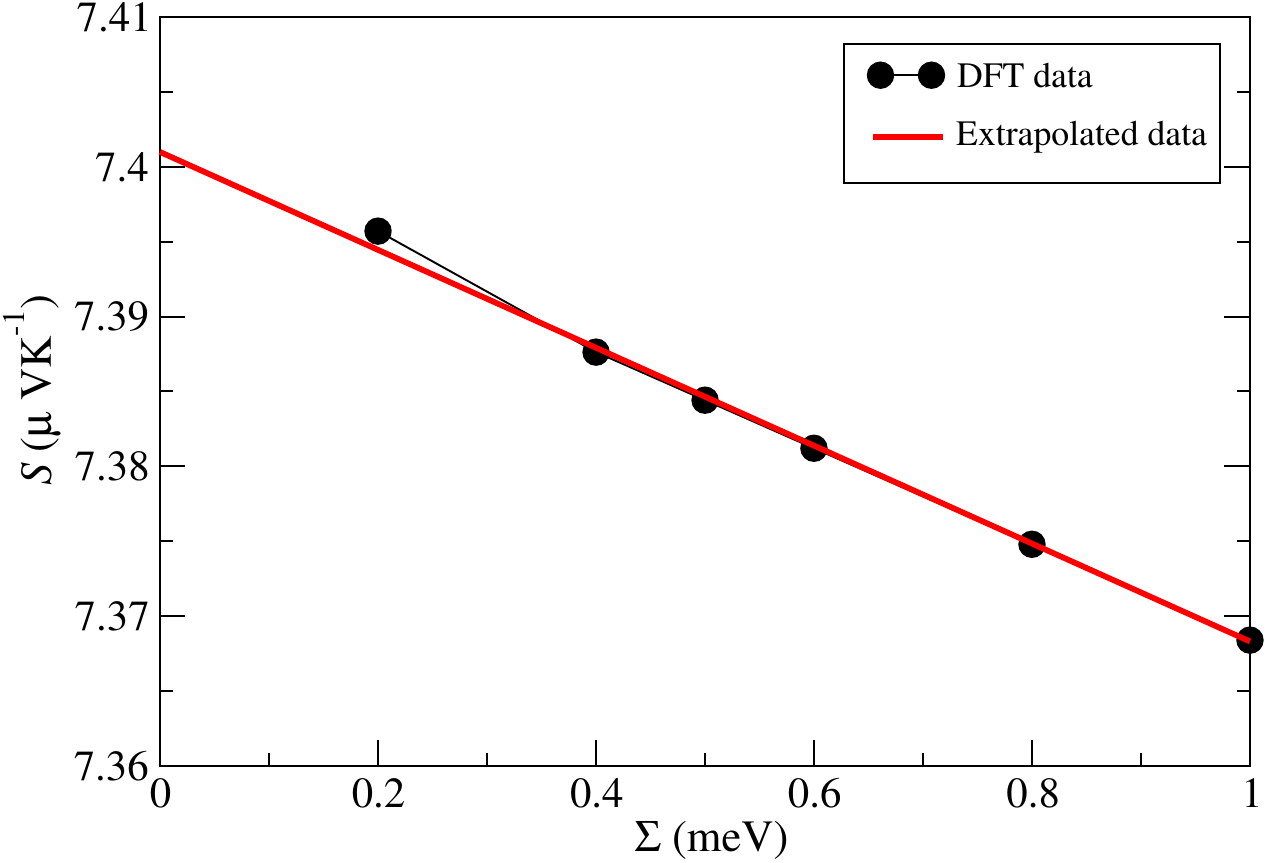}
\caption{\label{Fig.self_energy}\small{Imaginary part of self energy ($\Sigma$) dependent Seebeck coefficient ($S$) at 300 K. The black dots represent the calculated value of $S$ using TRACK$_{DFT}$ code\cite{sihi2021track}. The red line represents the extrapolated value of $S$ up to $\Sigma=0$ eV.}}
\end{figure}

The calculated value of $\kappa_e^{ee}$ is found to be very high in comparison to the experimental $\kappa$, which is shown in Fig. \ref{Fig.wth}. This may be due to the fact that in the present calculation, only EEIs are involved for which $\tau_{ee}$ is extremely high. However, in real, the materials possess all possible interactions. According to Matthiessen's rule, the combined effect of all the possible interactions reduces the value of $\tau$\cite{ashcroft}. Thus, it is expected that calculated values will be higher than experimental results. The results obtained in Fig. \ref{Fig.wth} are in accordance with this discussion. According to Wiedemann-Franz law\cite{ashcroft}, it is well known that $\kappa_e^{ee}$ depends on the $\rho_{ee}$ and $T$\cite{ashcroft}. As already discussed, the effect of electronic correlation on $\rho$ is extremely small. This suggests that the effect of electronic correlation on $\kappa$ will also be negligible. It is also important to note here that the calculated values of the $\kappa_{eph}$ is in good match with the available experimental data within the given temperature range\cite{Hellwege}. This generally suggests that EEIs are insignificant in deciding the values of $\kappa$. This further implies that electronic correlation is negligibly important to the values of $\kappa$ for V.

At any given value of $T$, $\kappa_e$ and $\rho$ are related to one another through Lorenz number $(L)$. This is popularly known as Wiedemann-Franz law\cite{ashcroft}. The calculation of $L$ for V in the given temperature range is carried out and corresponding results are mentioned in Table \ref{taba}. The obtained order of $L$ very well matches with the typical order of $L$ for the metallic system\cite{ashcroft}. Along with $L$, the effect of electronic correlation on the Seebeck coefficient ($S$) is also studied. 

\begin{figure}\label{Fig.band1} 
\centering
\includegraphics[width=0.49\linewidth, height=4.0cm]{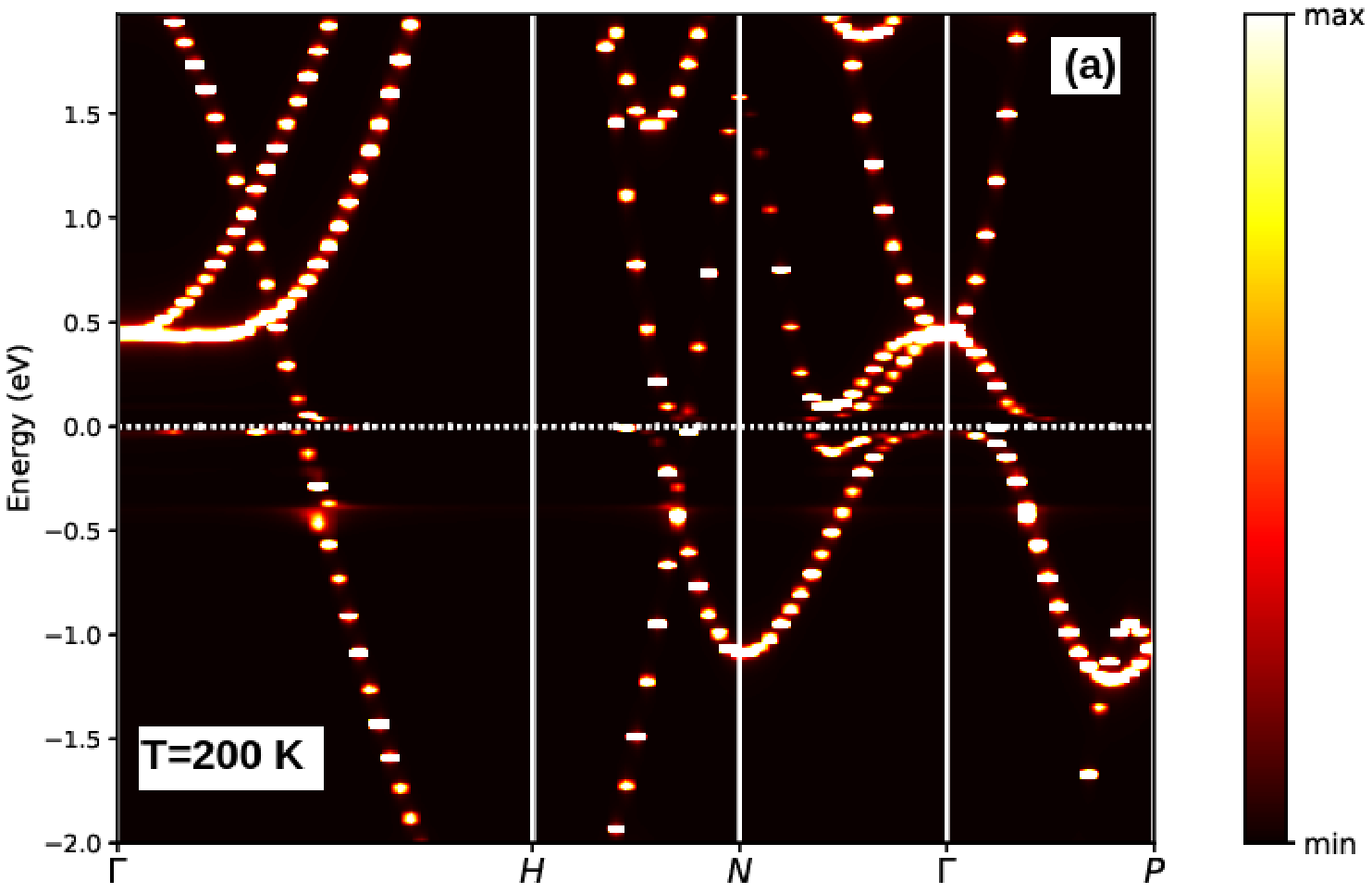}
\includegraphics[width=0.49\linewidth, height=4.0cm]{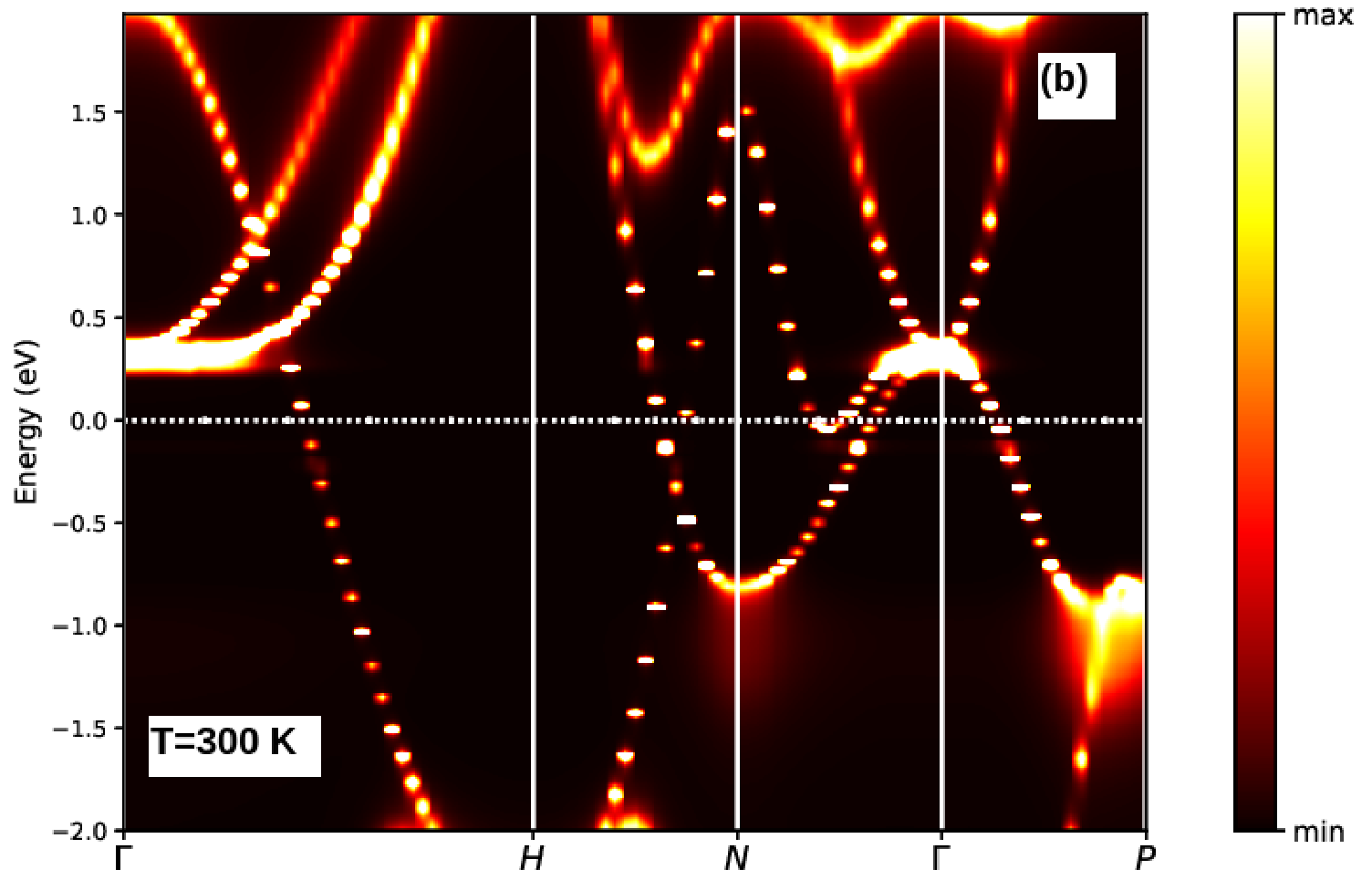}
\caption{\label{Fig.band1}\small{Momentum-resolved many-body spectral function (a) at
200 K (b) at 300 K. Fermi energy is scaled to 0 eV.}}
\end{figure}


The calculation of $S$ at 300 K using the DFT approach is carried out. In this direction, the values $S$ are computed at different magnitudes of $\Sigma$. Corresponding plot is shown in Fig. \ref{Fig.self_energy}. Using the extrapolation method, the value of $S$ has been obtained at $\Sigma=0$, which is equal to $\sim 7.401$ $\mu $VK$^{-1}$. This corresponds to real DFT value at 300 K. The obtained data for $S$ at DFT level is very off from the available experimental results\cite{Okram, Mackintosh}. Thus, it would be interesting to verify if the electronic correlation is reponsible for this inconsistency in DFT data with the experimental results. In light of this, the study of temperature-dependent $S$ is carried out at DFT+DMFT level within the temperature range $100-1000$ K. In these calculations, it is found that the values of $S$ are highly sensitive to the size of $k$-mesh used. So, to check the convergence of $S$ with respect to $k$-point, the calculations of $S$ at 300 K with different sizes of $k$-mesh are carried out. This is done to obtain an optimized size of $k$-mesh for studying the temperature-dependent behaviour of $S$ of the material. The values of $S$ corresponding to different sizes of $k$-mesh are mentioned in Table \ref{tabb}. For the larger size of $k$-mesh, it is found that there is a small fluctuation in the calculated value of $S$. At $300$ K, the values of $S$ are obtained as $1.4$, $-1.108$, $0.359$ and $-0.547$ $\mu $VK$^{-1}$ corresponding to the $k$-mesh size of ($200\times200\times200$), ($220\times220\times220$), ($250\times250\times250$) and ($300\times300\times300$), respectively. Taking the absolute values of $S$ corresponding to all four $k$-meshes yields the mean value of $0.8545$ $\mu $VK$^{-1}$. The obtained mean value is found to be in good agreement with the experimentally reported data\cite{Okram, Mackintosh} at 300 K with an error bar $\pm 0.2$ $\mu $VK$^{-1}$. Thus, for studying the temperature-dependent behaviour of $S$, optimized size of $k$-mesh is chosen as ($300\times300\times300$). Having found this, the calculations of temperature-dependent $S$ over the $k$-mesh size of ($300\times300\times300$) are carried out. The obtained results are mentioned in Table \ref{tabc}. It can be seen from table that for $T\le 300$ K, the values of $S$ are positive and its magnitude decreases with the rise in $T$. However, for $T>300$ K, the values of $S$ are negative and its magnitude increases with the rise in $T$. At 100 K, the value of $S$ is found to be $3.967$ $\mu $VK$^{-1}$ which gradually changes to $-42.96$ $\mu $VK$^{-1}$ at 1000 K. It is important to note that the results show the change in sign of $S$ as $T$ increases from 200 to 300 K.


\begin{table}\label{tabc}
\caption{\label{tabc}
\small{Calculated values of Seebeck coefficient through full type Coloumb interaction ($S_{full}$) at different temperatures ($T$) using $k$-mesh size of ($300\times 300\times 300$). Also, the experimental values and the reported theoretical values using Ising type Coloumb interaction ($S_{Ising}$) are mentioned.}} 
\begin{ruledtabular}
\begin{tabular}{lccccc}
\textrm{{$T$ (K)}}&
\textrm{{$S_{full}$ ($\mu $VK$^{-1}$})}&
\textrm{{$S_{experimental}$ ($\mu $VK$^{-1}$})}& 
\textrm{{$S_{Ising}$ ($\mu $VK$^{-1}$})}\\

\colrule
    100    & $3.967$ & 2.82\cite{Jung}, 1.02\cite{Okram} & -  \\
    200    & $4.39$  &  0.41\cite{Jung}, 0.27\cite{Okram}, 0.86\cite{Mackintosh}, 0.6\cite{Mackintosh} & 6.28\cite{sihi2021track} \\
    300    & $-0.547$ & -1.06\cite{Okram}, 1.0\cite{Mackintosh}  & 5.95\cite{sihi2021track} \\
    400    & $-14.48$ & -  & 1.84\cite{sihi2021track} \\
    500    & $-19.03$ & -  & - \\
    600    & $-26.81$ & -  & - \\
    700    & $-34.79$ & -  & - \\
    800    & $-42.82$ & -  & - \\
    900    & $-57.37$ & -  & - \\
    1000   & $-42.96$ & - &  -  \\
 
\end{tabular}
\end{ruledtabular}
\end{table}

From Table \ref{tabc}, it is found that for $T<300$ K, the values of $S$ are positive. Hence, below room temperature, this material must show $p$-type behaviour. However, at  higher values of $T$ ($\geq 400$ K), it is found that the sign of $S$ is negative. This shows that for values of $T\geq 400$ K, the material must exhibit $n$-type behaviour. From the table, at 100 and 200 K, it is also found that the values of $S$ is not in good match with the experimental results. The possible reason for this may be that the maximum entropy method is not efficient at lower values of $T$. Hence, for calculating $S$ at small values of $T$, one must use a dense $k$-mesh to obtain reasonable results. However, in the present work, due to higher computational cost, the calculations are not performed for a $k$-mesh size greater than ($300\times300\times300$). Furthermore, full type Coulomb interaction is used to calculate the values of $S$. Sihi \textit{et al.}\cite{sihi2021track} have also calculated the values of $S$ using Ising type of Coulomb interaction. But, in comparison to their data, the results of present work give better match with the experimental results, which is depicted in Table \ref{tabc}. This indicates that for V, full type Coulomb interaction is a better choice to calculate the values of $S$.

In order to understand the change in sign of $S$ between the values at 200 K and 300 K, the electronic dispersion curves are plotted at these temperatures, which are shown in Fig(s). \ref{Fig.band1}(a) and \ref{Fig.band1}(b), respectively. It is seen from Fig. \ref{Fig.band1}(a), near the Fermi level in the high symmetric directions $N-\Gamma-P$, the size of hole pocket is larger than the electron pocket. Thus, at 200 K, the concentration of holes is expected to be higher than the concentration of electrons at the Fermi level. In addition to this, it is seen that hole pocket is flatter in comparison to that of electron pocket. This generally suggests that the effective mass ($m^*$) for holes will be higher in comparison to that for electrons. In case of solids, $S$ is directly proportional to $m^*$ and inversely proportional to $n$\cite{Jeffrey}. The effect of $m^*$ seems to dominate over the effect of carrier concentration in deciding the value of $S$. This results in the positive value of $S$ at 200 K. Furthermore, in Fig. \ref{Fig.band1}(b), it is seen that near the Fermi level, there is only the presence of an electron pocket along $N-\Gamma$ direction. Thus, it is expected that value of $S$ should be negative, which is depicted in Table \ref{tabc}.

It is seen from the above discussion that shifting from DFT to the DFT+DMFT approach, the value of $S$ changes a lot. Moreover, unlike the DFT results, the value of $S$ at 300 K obtained using DFT+DMFT methods gives a very good match with the experimental data. This shows that electronic correlation plays a very important role in deciding the value of $S$ in case of V which is a moderately CESs.

\begin{figure}\label{Fig.BST1}
\centering
\includegraphics[width=0.90\linewidth, height=6.0cm]{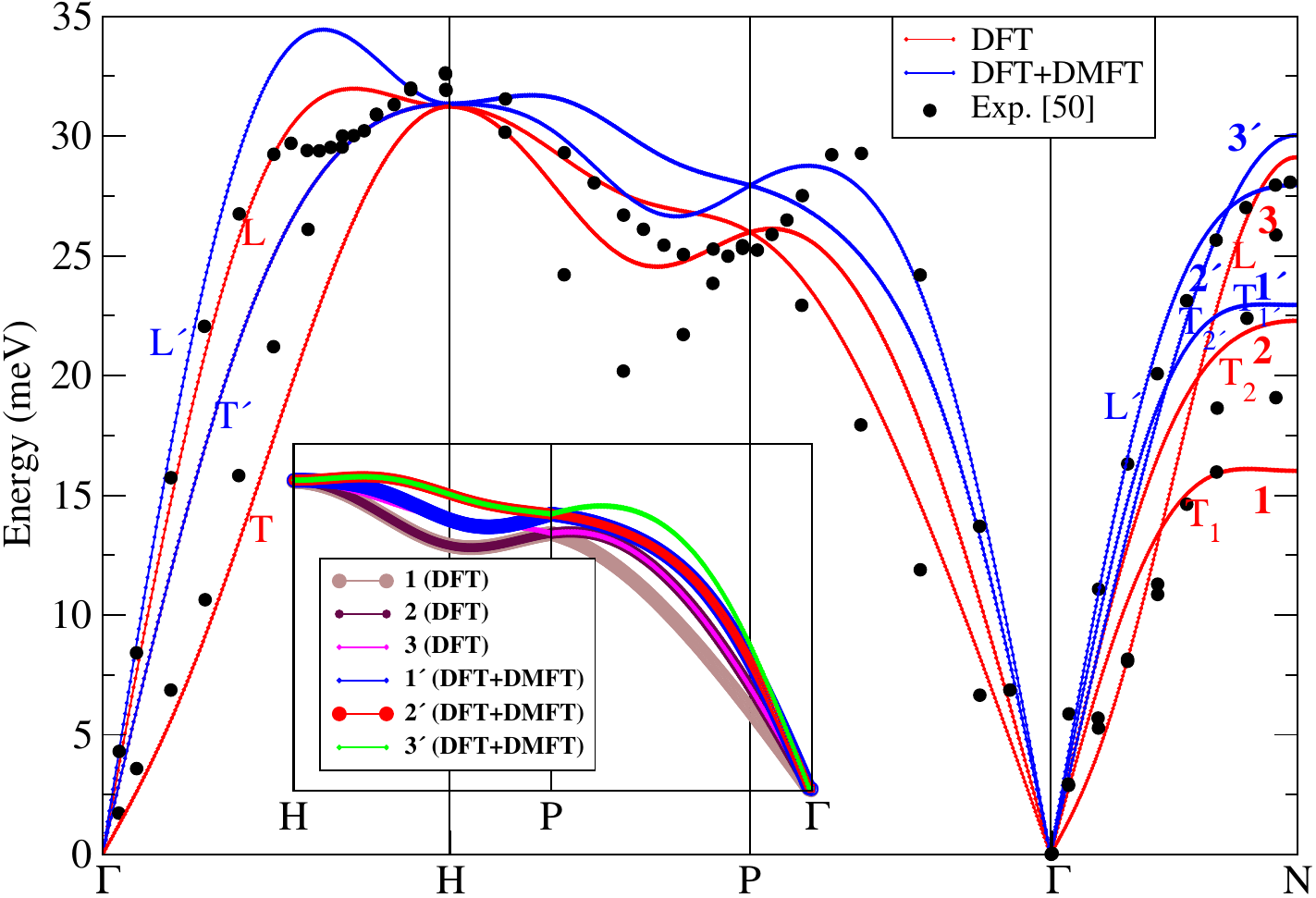}
\caption{\label{Fig.BST1}\small{Comparison between DFT, DFT+DMFT and experimental result\cite{Bosak} of phonon dispersion curves of V.}} 
\end{figure}

\subsection{\label{sec:level2} Phonon dispersion and phonon density of states}

The phonon dispersion curve obtained using DFT and DFT+DMFT based calculations are shown in Fig. \ref{Fig.BST1}. Also, the obtained results are compared with the available experimental data\cite{Bosak}. The curves from DFT+DMFT calculations are obtained at 300 K. These curves are plotted along the high symmetric directions $\Gamma-H-P-\Gamma-N$ in the first Brillouin zone. V is well-known to crystallize in a body-centered cubic crystal structure with one atom in the primitive cell. This is reflected in the phonon dispersion curve which exhibits three acoustic branches, here represented as 1 ($1^\prime$), 2 ($2^\prime$) and 3 ($3^\prime$) corresponding to DFT (DFT+DMFT). Among the three acoustic modes, two are transverse, (represented by T$_1$ (T$_{1^\prime}$) and T$_2$ (T$_{2^\prime}$)) and one is longitudinal (represented by L (L$^{\prime}$)). From the figure, it is seen that the bands at  $H$ and $P$ points are triply degenerate at both the levels. Nextly, along $\Gamma-H$ directions, the phonon bands 1 ($1^\prime$) and 2 ($2^\prime$) are degenerate while band 3 ($3^\prime$) is non-degenerate at DFT (DFT+DMFT) level. Furthermore, along $H-P$ directions, bands 1 and 2 are degenerate while band 3 is non-degenerate at DFT level which is shown in subplot of Fig. \ref{Fig.BST1}. But, at DFT+DMFT level, along $H-P$ directions, bands $2^\prime$ and $3^\prime$ are degenerate while band $1^\prime$ is non-degenerate. Moreover, at DFT level, it is further seen that along $P-\Gamma$ directions, bands 2 and 3 are degenerate while band 1 is non-degenerate. However, at DFT+DMFT level, bands $1^\prime$ and $2^\prime$ are degenerate and band $3^\prime$ is non-degenerate along $P-\Gamma$ direction. In addition to this, along $\Gamma-N$ directions all the three bands are found to be non-degenerate at both the levels. Moving further, in most of the regions along the given high symmetric directions, the three phonon bands obtained at DFT+DMFT level have higher energy than the corresponding bands obtained from DFT calculations. However, at $H$ high symmetric point, the calculated bands have same energy at both the levels. Along $\Gamma-N$ directions, the energy of band 1 obtained from DFT+DMFT calculations is greater than the corresponding energy obtained at DFT level. The similar behaviour is seen for bands 2 and 3. Apart from this, it is also seen that at $N$ point, the gap between bands 2 and 3 is greater in DFT as compared to DFT+DMFT calculations.

The phonon dispersion curve obtained from DFT and DFT+DMFT differ from each other to a good extent. The better agreement between DFT+DMFT results with available experimental data\cite{Bosak} of the phonon dispersion curve is strong evidence that electronic correlation is extremely importance in studying the phononic states of V. 

\begin{figure} \label{Fig.DOS}
\centering
\includegraphics[width=0.90\linewidth, height=6.0cm]{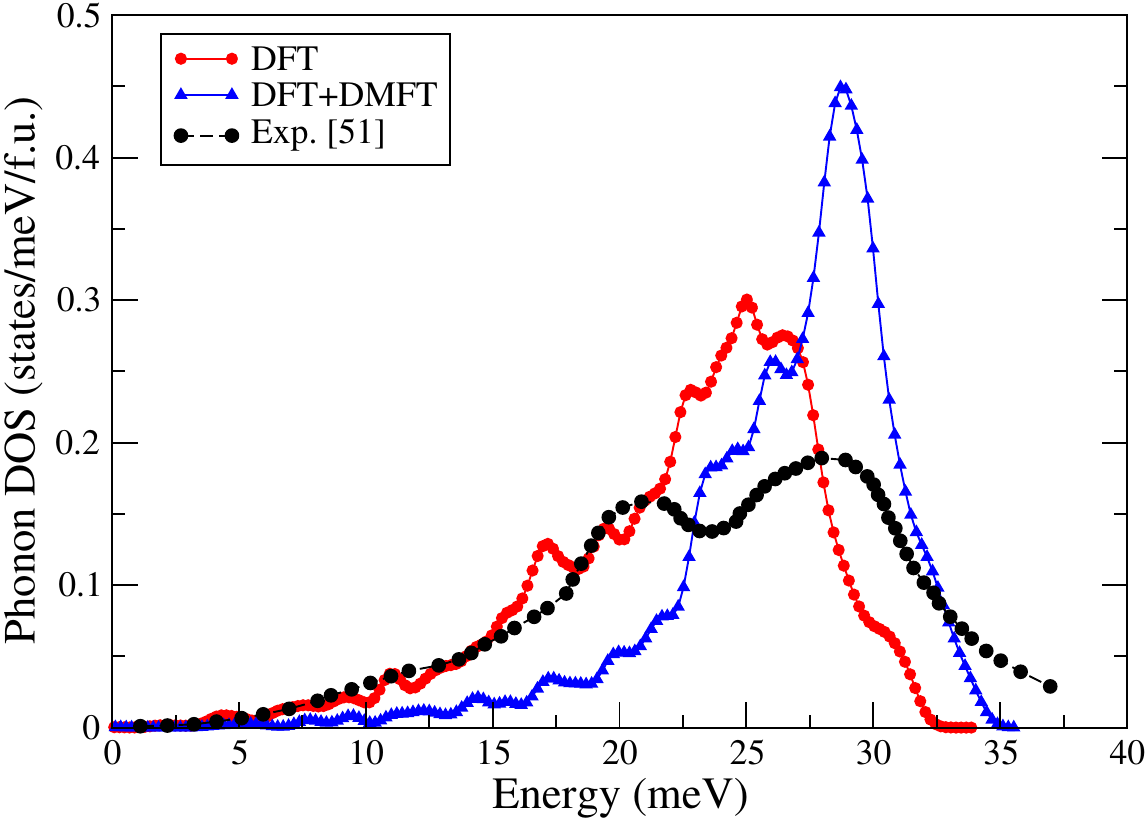}
\caption{\label{Fig.DOS}\small{ Comparison between DFT, DFT+DMFT and experimental data\cite{Eisenhauer} of phonon density of states (DOS) of V.}}
\end{figure} 

In order to have a better picture of electronic correlation on phononic states, the phonon DOS are also calculated using both the methods i.e., DFT and DFT+DMFT. The DOS at DFT+DMFT level are calculated at 300 K. Corresponding plots are shown in Fig. \ref{Fig.DOS} along with the available experimental result\cite{Eisenhauer}. The maximum value of DOS is found to be $\sim$ 0.30 at the energy $\sim$ 25 meV using DFT calculations. This energy, in terms of temperature, corresponds to $\sim$ 290 K. In the case of DFT+DMFT computations, the maximum value of DOS is found to be $\sim$ 0.45 at the energy $\sim$ 28.70 meV, which corresponds to $\sim$ 333 K. Furthermore, the maximum phonon energy above which DOS is zero is $\sim$ 33.83 meV at DFT level. At DFT+DMFT level, this energy corresponds to $\sim$ 35.15 meV. This value is more closer to the reported data of maximum phonon energy at room temperature in various experimental works, which are 37.22\cite{Eisenhauer, Turberfield}, 35.15\cite{Stewart}, 36.98\cite{Mozer} and 41.77\cite{chernoplekov1963investigation}meV. From the figure, it is seen that the deviations between the calculated values of phonon DOS at the DFT level and experimental results\cite{Stewart, Eisenhauer, Turberfield} are large. While the DFT+DMFT approach improves the deviations above $\sim 28$ meV. It is also seen that the maximum value of DOS calculated at the DFT+DMFT calculation is in a good agreement with the experimental result\cite{Eisenhauer}. Thus, this study shows that the electronic correlation has a tremendous effect on phononic states, even in moderately CESs like V.

\begin{table}\label{tabd}
\caption{\label{tabd}
\small{Comparison of the present results with available experimental data of elastic constants and bulk modulus ($B$) (units in GPa). The experimental values are at room temperature.}} 
\begin{ruledtabular}
\begin{tabular}{lccccc}
\textrm{{}}&
\textrm{{$C_{44}$ }}&
\textrm{$C_{11}$ }& 
\textrm{{$C_{12}$}}& 
\textrm{{$B$}}\\ 
      
\colrule
    DFT                           & 41.49 & 223.06 & 120.87 & 154.93\\
    DFT+DMFT                      & 48.87 & 228.67 & 111.88 & 150.8\\
    Exp.\cite{PhysRev.119.1532}   & 42.55 & 227.95 & 118.7 & 155.1\\
    Exp.\cite{10.1063/1.1735933}  & 42.6  & 228    & 119 & 155.5\\
 
\end{tabular}
\end{ruledtabular}
\end{table}

Along with phonon dispersion and phonon DOS, the elastic constants and  bulk modulus is also studied. The slopes of longitudinal and transversal phonon branches near the $\Gamma$ point provides the value of the elastic constants ($C_{44}$, $C_{11}$ and $C_{12}$). The calculations of elastic constants and bulk modulus ($B$) are performed using the relations as follows\cite{PhysRevB.2.4176}: 

\begin{equation}\label{eq.V}
v=\frac{\mathrm{d}\omega}{\mathrm{d}|\vec{q}|}
\end{equation}

\begin{equation}\label{eq.C11}
\rho v_{1}^2=C_{11}
\end{equation}

\begin{equation}\label{eq.C44}
\rho v_{2}^2=C_{44}
\end{equation}

\begin{equation}\label{eq.C12}
\rho v_{3}^2=\frac{1}{2}(C_{11}-C_{12})
\end{equation}
and
\begin{equation}\label{eq.B}
B=\frac{1}{3}(C_{11}+2C_{12})
\end{equation}

where $v$, $\omega$ and $|\vec{q}|$ represent the wave velocity, frequency and wave vector, respectively. $\rho$ denotes the mass density of the atom. Also, $v_1$, $v_2$ and $v_3$ correspond to longitudinal (100), transverse (100) and slow transverse (110) waves, respectively. The values of elastic constants corresponding to $C_{44}$, $C_{11}$ and $C_{12}$ are mentioned in Table \ref{tabd}. From the table, it can be seen that the value of $C_{44}$, $C_{11}$ and $C_{12}$ are 41.49 (48.87), 223.06 (228.67) and 120.87 (111.88) GPa, respectively at DFT (DFT+DMFT) level. The obtained elastic constants are found to be in a good agreement with the available experimental data\cite{PhysRev.119.1532, 10.1063/1.1735933}. The bulk modulus $B$ is also calculated and mentioned in Table \ref{tabd}. The obtained value of $B$ is found to be 154.93 (150.8) GPa at DFT (DFT+DMFT) level, which is in a good match with the experimental data\cite{PhysRev.119.1532, 10.1063/1.1735933}.

\section{Conclusions}
The study of transport properties shows that the values of $\rho$ due to EEIs are negligible in the comparison to $\rho$ due to EPIs within the temperature range $100-1000$ K. In addition to this, the value of $\kappa_e$ due to EPIs is found to be in good agreement with the available experimental data at any given $T$, within $20-300$ K. These observations generally suggest that EEIs play insignificant role in deciding the values of these properties within the studied temperature range. Moving further, at 300 K, the values of $S$ obtained at DFT level $\sim 7.401$ $\mu$VK$^{-1}$. The value of $S$ at 300 K, calculated using DFT+DMFT method, is found to be $\sim$ -0.547$\mu$VK$^{-1}$, which is very much off from DFT results. Apart from the transport properties, the phonon dispersion curve calculated at DFT+DMFT level is found to differ a lot from that at DFT level. Furthermore, phononic DOS obtained at DFT+DMFT level shows considerable differences from the DOS at DFT level. The maximum energy of phononic states obtained at DFT and DFT+DMFT level are $\sim$ 33.83 and $\sim$ 35.15 meV, respectively. These observations clearly indicate that electronic correlation is considerably important in studying the $S$ \& phononic states of V and must not be neglected.

\bibliography{MS}
\bibliographystyle{apsrev4-2}

\end{document}